# Review of a light NMSSM pseudoscalar Higgs state production at the LHC

M. M. Almarashi[*1]

*Department of Physics, Faculty of Science,
Taibah University, P.O.Box 30002, Madinah, Saudi Arabia

**Abstract**

In this paper we briefly review the LHC discovery potential of a light pseudoscalar Higgs boson of the NMSSM, $a_1$, produced in the gluon fusion $gg \to a_1$, bottom-quark fusion $b\bar{b} \to a_1$ and bottom-gluon fusion $bg \to ba_1$. We also review the LHC discovery potential of the next-to-lightest CP-even Higgs boson $h_2$ being the non-SM-like Higgs, decaying either into two light CP-odd Higgs bosons $a_1 a_1$ or into a light $a_1$ and the $Z$ gauge boson through the gluon fusion $gg \to h_2$ in the $4\tau$ final state. We find that the light $a_1$ can be detected at the LHC in a variety of production processes including the gluon fusion, bottom-quark fusion and bottom-gluon fusion. The latter two processes require high luminosity of the LHC and large values of $\tan\beta$. We also find that the LHC has the potential to discover the non-SM-like Higgs state, $h_2$, decaying into a pair of light CP-odd Higgses $a_1$'s, allowing to distinguish the NMSSM Higgs sector from the MSSM one as such a light $a_1$ is impossible in the latter scenario.

---

[1] mmarashi@taibahu.edu.sa